\documentclass[twocolumn,showpacs,pr,aps]{revtex4-1}
\usepackage{bm}
\usepackage{graphicx}
\usepackage{epsfig}
\usepackage{amsmath} 
\usepackage{color}
\usepackage{ulem}
\usepackage{amssymb}
\usepackage[colorlinks = true]{hyperref}

\renewcommand{\i}{\mathrm{i}}
\newcommand{\eps}{\varepsilon}
\newcommand{\e}{\mathrm{e}}

\newcommand{\dd}[2]{\frac{\partial #1}{\partial #2}}

\newcommand{\kp}{$\bm{k} \cdot \bm{p}$}

\definecolor{darkgray}{rgb}{0.66, 0.66, 0.66}

\begin{document}

\title{Atomically inspired $k \cdot p$~approach and valley Zeeman effect in transition metal dichalcogenide monolayers}

\author{D. V. Rybkovskiy}
\affiliation{A. M. Prokhorov General Physics Institute, RAS, 38 Vavilov street, 119991, Moscow, Russia}
\affiliation{Faculty of Physics, Southern Federal University, 5 Zorge Street, 344090, Rostov-on-Don, Russia}
\author{I. C. Gerber}
\affiliation{Universit\'e de Toulouse, INSA-CNRS-UPS, LPCNO, 135 Ave. de Rangueil, 31077 Toulouse, France}
\author{M. V. Durnev}
\email[correspondence address: ]{durnev@mail.ioffe.ru}
\affiliation{Ioffe Institute, 194021 St.\,Petersburg, Russia}

\begin{abstract}
We developed a six-band \kp~model that describes the electronic states of monolayer transition metal dichalcogenides (TMDCs) in $\bm K$-valleys. The set of parameters for the \kp~model is uniquely determined
 by decomposing tight-binding (TB) models in the vicinity of $\bm K^\pm$-points. First, we used TB models existing in literature  to derive systematic parametrizations for different materials, including MoS$_2$, WS$_2$, MoSe$_2$ and WSe$_2$.  Then, by using the derived six-band \kp~Hamiltonian we calculated effective masses, Landau levels, and the effective exciton $g$-factor $g_{X^0}$ in different TMDCs. 
We showed that TB parameterizations existing in literature result in small absolute values of $g_{X^0}$, which are far from the experimentally measured $g_{X^0} \approx -4$. To further investigate this issue we derived two additional sets of \kp~parameters by developing our own TB parameterizations based on simultaneous fitting of ab-initio calculated, within the density functional (DFT) and $GW$ approaches, energy dispersion and the value of $g_{X^0}$. We showed that the change in TB parameters, which only slightly affects the dispersion of higher conduction and deep valence bands, may result in a significant increase of $|g_{X^0}|$, yielding close-to-experiment values of $g_{X^0}$. Such a high parameter sensitivity of $g_{X^0}$ opens a way to further improvement of DFT and TB models.

\end{abstract}

\maketitle 

\section{Introduction}

Monolayers of transition metal dichalcogenides (TMDCs) have attracted recently much attention due to their exceptional properties, such as coupling of spin and valley degrees of freedom, which allows for the valley polarization with a circularly polarized light in these materials~\cite{PhysRevLett.105.136805, Wang:2012rr, PhysRevLett.108.196802, Mak:2012oq, Cao:2012ai, PhysRevB.86.081301}. Recent magneto-photoluminescence experiments revealed the significant Zeeman splitting of emission lines associated with optical recombination of excitons and trions in different valleys. The effective exciton $g$-factors of this ``valley Zeeman effect'' in a magnetic field directed normal to a monolayer plane were found to be $g_{X^0} \approx -4$ for a wide range of investigated materials, including selenides~\cite{PhysRevLett.113.266804, PhysRevLett.114.037401, Srivastava:2015db, Aivazian:2015zl, wang2015, Mitioglu:2015ij}, sulphides~\cite{Stier:2016dk, PhysRevB.93.165412} and tellurides~\cite{Arora:2016lk}.

The multi band \kp~method is perfectly suited for theoretical investigation of magneto-optical and transport effects, including Zeeman effect, optical absorption and photogalvanics~\cite{ivchenkopikus, voon}. The available two-band \kp~models describing electronic spectra in TMDCs~\cite{kormanyos2015,PhysRevX.4.011034,PhysRevB.88.045416} account for the bottom conduction and topmost valence bands and are parameterized by density functional theory (DFT) calculations. The simple two-band model is, however, insufficient for calculation of exciton Zeeman effect since the exciton $g$-factor $g_{X^0}$ vanishes in the two-band \kp~approximation~[\onlinecite{wang2015}]. Therefore to obtain nonzero $g_{X^0}$ additional terms describing contributions of remote bands should be included in the two-band \kp~model~\cite{PhysRevX.4.011034, PhysRevB.88.085440}, which make the model less transparent. Moreover parametrization of the \kp~model by fitting the DFT band structure only in the vicinity of distinct points of the Brillouin zone is less reliable and allows for a much freedom in the choice of parameters.


In this work we use an alternative approach based on the idea proposed in Ref.~[\onlinecite{wang2015}]. This approach comprises three steps, namely, (i) use DFT-based calculations of electronic states in TMDCs as a starting point, (ii) use atomistic tight-binding (TB) model to fit the electronic spectrum and wave functions along the high-symmetry paths of the Brillouin zone, and (iii) derive \kp~Hamiltonian by decomposing the TB model in the vicinity of a given wave vector in the Brillouin zone. As a result, we obtain a multi band \kp~Hamiltonian (the number of bands is equal to the number of atomic orbitals included in the TB model) with a set of parameters that are uniquely determined by the TB parametrization. This Hamiltonian can be further used for calculating the valley Zeeman effect, Landau levels, etc.

As a starting point we use several existing eleven-band TB models~\cite{PhysRevB.88.075409, 2053-1583-1-3-034003, PhysRevB.92.205108, 0953-8984-27-36-365501, PhysRevB.92.195402}.
These models include $d$-type orbitals of metal atoms and $p$-type orbitals of chalcogen atoms, and capture all symmetries of the studied system. The resulting \kp~Hamiltonian, which describes dispersion of the bottom conduction and topmost valence bands, contains six bands that have even parity with respect to mirror reflection in the monolayer plane. We present a systematic parametrization of the six-band \kp~Hamiltonian by expansion of TB Hamiltonians~\cite{PhysRevB.88.075409, 2053-1583-1-3-034003, PhysRevB.92.205108, 0953-8984-27-36-365501, PhysRevB.92.195402} in the vicinity of $\bm K^\pm$ points of the Brillouin zone for different materials (MoS$_2$, WS$_2$, MoSe$_2$, WSe$_2$) and different TB models. We also do the whole three-step procedure on our own, i.e., perform post-DFT calculations by applying $GW$ formalism to obtain reliable band gaps and accurate band dispersion, fit it with the TB model and do the \kp~decomposition, which allows us, as a result, to obtain an independent \kp~parameterization. We want to stress the novelty of using $GW$ set of data to extract TB parameters since in previous studies the main focus was made on DFT calculations using mainly local or semi-local exchange-correlation functionals~\cite{PhysRevB.88.075409, 2053-1583-1-3-034003, PhysRevB.92.205108}, whereas hybrid functional was also used~\cite{0953-8984-27-36-365501} to partially correct the severe underestimation of band gap values usually observed at the DFT level for MoS$_2$ monolayers, see Ref.~[\onlinecite{MolinaSanchez:2015fs}] for a recent review. The use of $GW$ correction strongly affects the effective charge carrier masses~\cite{MolinaSanchez:2013hz,Shi:2013jv} too.   

Then, we use the derived six-band \kp~Hamiltonian to calculate effective masses, Landau levels, and the effective exciton $g$-factor $g_{X^0}$ in different TMDCs. We analyze the main contributions to $g_{X^0}$, which result from the mixing with excited conduction and deep valence bands. We show that the calculated values of $g_{X^0}$ and effective masses vary in a wide range for different TB parameterizations. The calculated values of $g_{X^0}$ for available in literature TB models ($|g_{X^0}| \lesssim 1$) are rather far from experimental values. However using our DFT+$GW$ calculations and the TB fitting procedure we were able to find additional \kp~parameterization sets, which well describe the experimental exciton $g$-factor ($g_{X^0} \approx -4$), as well as conduction and valence band effective masses and the wave functions coefficients. This result underlies the importance of the careful choice for the DFT starting point calculations and TB parameterizations: along with effective masses and energy gaps the $g$-factor value may serve as a test for improving both the DFT calculations and TB models.

\section{Effective kp-Hamiltonian}

As a starting point for construction of an effective \kp-Hamiltonian we will use eleven-band tight-binding models developed in Refs.~\onlinecite{PhysRevB.88.075409, 2053-1583-1-3-034003, PhysRevB.92.205108, 0953-8984-27-36-365501, PhysRevB.92.195402}. These tight-binding models include three $p$-type orbitals on each of the two chalcogen atoms (X) and five $d$-type orbitals on a metal atom (M). The electron wave function within the tight-binding approximation is presented as a linear combination of atomic orbitals $\phi_j^{(a)}$~\cite{wang2015}
\begin{equation}
\label{eq:wf_TB}
\Psi_{\bm k}^{(n)} (\bm r) = \sum \limits_{a,l,j} \e^{\i \bm k \bm R_{a,l}} C_{j}^{(a)} \phi_j^{(a)} (\bm r - \bm R_{a,l})\:,
\end{equation}
where $n$ is a number of an electronic band, $\bm k$ is a wave vector, $a = \mathrm M, \mathrm X$ denotes the type of an atom, $l$ runs through the atoms of a given type, $j$ enumerates the set of orbitals, $\bm R_{a,l}$ gives the position of atoms in a two-dimensional lattice, and $C_{j}^{(a)}$ are coefficients.

The basis orbitals of the eleven-band tight-binding model are~\cite{PhysRevB.88.075409, 2053-1583-1-3-034003, PhysRevB.92.205108, 0953-8984-27-36-365501, PhysRevB.92.195402}
\begin{multline}
\label{eq:TB_basis} 
\phi_j^{(a)} = 
\left \{ d_{z^2}, d_+, d_-, p_+, p_-, p_{z,A}, \right.
\\ \left. d_{xz}, d_{yz}, p_{x, A}, p_{y, A}, p_{z,S} \right\}\:,
\end{multline}
where 
$d_\pm = d_{x^2 - y^2} \pm 2\i d_{xy}$, $p_\pm = p_{x,S} \pm \i p_{y,S}$, 
$d_\alpha$ denotes the orbital with a $d$-like symmetry of the M atom and $p_{\beta, S} = (p_{\beta, t} + p_{\beta, b})/\sqrt{2}$, $p_{\beta, A} = (p_{\beta, t} - p_{\beta, b})/\sqrt{2}$ are the symmetric and asymmetric combinations of the $p$-type orbitals of the top ($t$) and bottom ($b$) X atoms in the unit cell, $x$ and $y$ axes lie in the monolayer plane, and $z$ is the monolayer normal. The spin orbit interaction between electron spin and orbital momenta of atomic orbitals~\cite{2053-1583-1-3-034003, PhysRevB.92.205108} is neglected in this work.

If the $z \to -z$ mirror symmetry is conserved, i.e. for a free-standing monolayer in the absence of external electric field and strain, the Hamiltonian, which describes the energy spectrum of a monolayer electron with a wave vector $\bm k$, written in the basis Eq.~\eqref{eq:TB_basis} has the form
\begin{equation}
\mathcal H (\bm k) = 
\left(
\begin{array}{cc}
\mathcal H_E & 0 \\
0 & \mathcal H_O
\end{array}
\right)\:.
\end{equation}
Here $\mathcal H_E$ is the 6$\times$6 block acting on the orbitals with even with respect to $z \to -z$ symmetry, and $\mathcal H_O$ is the 5$\times$5 block acting on the orbitals with odd symmetry. The exact form of blocks $\mathcal H_E$ and $\mathcal H_O$ depends on a particular tight-binding model~\cite{PhysRevB.88.075409, 2053-1583-1-3-034003, PhysRevB.92.205108, 0953-8984-27-36-365501, PhysRevB.92.195402}. It is known that the Bloch functions of the bottom conduction and topmost valence bands, which are of the main interest in this work, are even with respect to $z \to -z$ reflection~\cite{PhysRevB.88.045416, kormanyos2015}, and therefore these bands are described by the $\mathcal H_E$ block. We note that magnetic field normal to a monolayer does not break the parity of wave functions, and hence we do not need the $\mathcal H_O$ block in the calculation of $g_{X^0}$. We also note, that the mixing of the $\mathcal H_E$ and $\mathcal H_O$ blocks by a perturbation that breaks $z \to -z$ symmetry does not affect $g_{X^0}$ in the first order in this perturbation.

To construct an effective \kp~Hamiltonian in the vicinity of $\bm K^\pm = (\pm 4\pi/3a_0,0)$ points, where $a_0$ is the lattice constant, we will decompose the tight-binding Hamiltonian $\mathcal H_E (\bm k)$ over a small wave vector $\bm q = \bm k - \bm K^{\pm}$. Up to the second-order terms this decomposition yields
\begin{multline}
\label{eq:decompose}
\mathcal H_{E}^{\pm} (\bm q) \approx \mathcal H_E (\bm K^\pm) + \sum \limits_{\alpha = x,y} \dd{\mathcal H_E}{k_\alpha} (\bm K^\pm) q_\alpha + \\
+ \frac12 \sum \limits_{\alpha, \beta = x,y} \frac{\partial^2 \mathcal H_E}{\partial k_\alpha \partial k_\beta} (\bm K^\pm) q_\alpha q_\beta\:.
\end{multline}

Electron wave functions at $\bm K^\pm$ valleys of MX$_2$ transform according to irreducible representations (irreps) of the C$_{3h}$ point group. We denote six wave functions that diagonalize $\mathcal H_E^+(\bm q)$ at $\bm q = 0$ as $\Psi_{E'_1}^{(v-5)}$, $\Psi_{A'}^{(v-4)}$, $\Psi_{E'_2}^{(v-3)}$, $\Psi_{A'}^{(v)}$, $\Psi_{E'_1}^{(c)}$, $\Psi_{E'_2}^{(c+2)}$, where a superscript names the electronic band and a subscript denotes the corresponding irreducible representation ($A'$, $E'_1$ and $E'_2$), see Tab.~\ref{tab:table1}.
 Note that in addition to two conduction ($c$ and $c+2$) and two valence ($v$ and $v-3$) bands, known from the four-band \kp~models~\cite{PhysRevB.88.045416, kormanyos2015}, the six-band model contains two deep valence bands $v-4$ and $v-5$, which transform at $\bm K^+$-point via $A'$ and $E'_1$ representations, respectively.

In the new basis $\mathcal H_E^\pm(\bm q)$ can be written as
\begin{equation}
\label{eq:Hkp}
\mathcal H_E^{\pm} (\bm q) = \mathcal H_1^{\pm} (\bm q) + \mathcal H_2^{\pm} (\bm q)\:,
\end{equation}
where the first term contains linear in $\bm q$ terms:
\begin{equation}
\label{eq:H1}
\mathcal H_1^+ (\bm q) = \left(
\begin{array}{cccccc}
E_{v-5} & \delta_7 q_- &\delta_6 q_+ & \delta_4 q_- & 0 & \delta_2 q_+ \\
\delta_7 q_+ & E_{v-4} & \delta_5 q_- & 0 & \delta_3 q_+ & \delta_1 q_- \\
\delta_6 q_- & \delta_5 q_+ & E_{v-3} & \gamma_2 q_+ & \gamma_5 q_- & 0 \\
\delta_4 q_+ & 0 & \gamma_2 q_- & E_{v}  & \gamma_3 q_+ & \gamma_4 q_- \\
0 & \delta_3 q_- & \gamma_5 q_+ & \gamma_3 q_- & E_{c}  & \gamma_6 q_+ \\
\delta_2 q_- & \delta_1 q_+ & 0 & \gamma_4 q_+ &\gamma_6 q_- & E_{c+2}\\
\end{array}
\right)\:,
\end{equation}
and the second one 
\begin{equation}
\label{eq:H2}
\left[\mathcal H_2^\pm (\bm q) \right]_{nl} = \frac{\hbar^2 q^2}{2 m'_{n}} \delta_{nl}\:,\:\: n,l = 1..6
\end{equation}
is a diagonal matrix with quadratic in $\bm q$ elements. In Eqs.~\eqref{eq:H1}, \eqref{eq:H2} $E_n$ ($n = c+2, c\dots$) are the band energies at $\bm K^\pm$-points, $\gamma_j$ and $\delta_j$ are parameters, $q_\pm = q_x \pm \i q_y$, and $q^2 = q_x^2 + q_y^2$. The effective masses $m'_n$ describe contributions to the band dispersion arising from the mixing with remote bands, which are not present in the \kp~model~\cite{wang2015}. Note that in the decomposition of the off-diagonal elements of the tight-binding Hamiltonian~\eqref{eq:decompose} we retained only the linear in $\bm q$ terms. The phases of wave functions at $\bm K^+$-point (Tab.~\ref{tab:table1}) are chosen in such a way that parameters $\gamma_j$ and $\delta_j$ are real.
Parameters of the \kp~Hamiltonians~\eqref{eq:H1} and \eqref{eq:H2} for different parametrizations, materials and tight-binding models are listed in Tab.~\ref{tab:table3} of the Appendix~\ref{app:kp}.

To derive the \kp~Hamiltonian $\mathcal H_1^-$ at the $\bm K^-$-point of the Brillouin zone one should replace $q_+$ by $q_-$ and vice versa in Eq.~\eqref{eq:H1}~\footnote{Note that this rule depends on the choice of the wave functions phase, i.e. in Ref.~\cite{kormanyos2015} $q_\pm \to - q_\mp$}. Note that at $\bm k = \bm K^-$ basis wave functions $\Psi^{(c)}$ and $\Psi^{(v-5)}$ transform according to $E'_2$ irreducible representation whereas $\Psi^{(c+2)}$ and $\Psi^{(v-3)}$ transform according to $E'_1$.


\begin{table}[b]
\caption{\label{tab:table1} The nonzero coefficients $C_j^{(a)}$ of wave functions Eq.~\eqref{eq:wf_TB} at $\bm K^+$-point and corresponding irreducible representations of the $C_{3h}$ point group. The phases of the wave functions are chosen in such a way, that $\alpha_i$ and $\beta_i$ are real numbers, $\alpha_i^2 + \beta_i^2 = 1$, and $\alpha_i > 0$. We denote $p_z \equiv p_{z,A}$.
}
\begin{ruledtabular}
\begin{tabular}{c|c|c}
Irrep & Band & Nonzero wave function coefficients \\
\hline
$A^\prime$ & $v$, $v-4$ & $\Psi^{(v)}$:  $C_{d_+} = \alpha_1$, $C_{p_+} = \i \beta_1$; \\
 & & $\Psi^{(v-4)}$: $C_{d_+} = \beta_1$, $C_{p_+} = -\i \alpha_1$ \\
 \hline
 $E^\prime_1$ & $c$, $v-5$ & $\Psi^{(c)}$: $C_{d_{z^2}} = \alpha_2$, $C_{p_-} = \i \beta_2$; \\
 & & $\Psi^{(v-5)}$: $C_{d_{z^2}} = \beta_2$, $C_{p_-} = -\i \alpha_2$ \\
 \hline
 $E^\prime_2$ & $c+2$, $v-3$ & $\Psi^{(c+2)}$: $C_{d_-} = \alpha_3$, $C_{p_{z}} = \beta_3$; \\
 & & $\Psi^{(v-3)}$: $C_{d_-} = \beta_3$, $C_{p_{z}} = -\alpha_3$
\end{tabular}
\end{ruledtabular}
\end{table}

The effective masses of the main conduction and valence bands in the framework of \kp~model are
\begin{equation}
\label{eq:masses}
\frac{1}{m_c} = \frac{1}{m'_c} + \frac{1}{m^*_c}\:,\:\:\: \frac{1}{m_v} = \frac{1}{m'_v} + \frac{1}{m^*_v}\:,
\end{equation}
where
\begin{align}
\label{eq:renormc}
\frac{1}{m^*_c} =  \frac{2}{\hbar^2}
& \left( \frac{\gamma_5^2}{E_c - E_{v-3}} + \frac{\gamma_3^2}{E_c - E_v} + \right. \\
&\left. + \frac{\gamma_6^2}{E_c - E_{c+2}} + \frac{\delta_3^2}{E_c - E_{v-4}} \right)\:, \nonumber \\
\label{eq:renormv}
\frac{1}{m^*_v} =  \frac{2}{ \hbar^2}
& \left( \frac{\gamma_2^2}{E_v - E_{v-3}} + \frac{\gamma_3^2}{E_v - E_c} + \right. \\
&\left. + \frac{\gamma_4^2}{E_v - E_{c+2}} + \frac{\delta_4^2}{E_v - E_{v-5}} \right) \nonumber
\end{align}
result from the mixing of electronic bands described by Hamiltonian~\eqref{eq:H1}, whereas $m'_c$ and $m'_v$ account for the mixing with remote bands, see Eq.~\eqref{eq:H2}. 

\begin{figure}[htpb]
\includegraphics[width=0.99\linewidth]{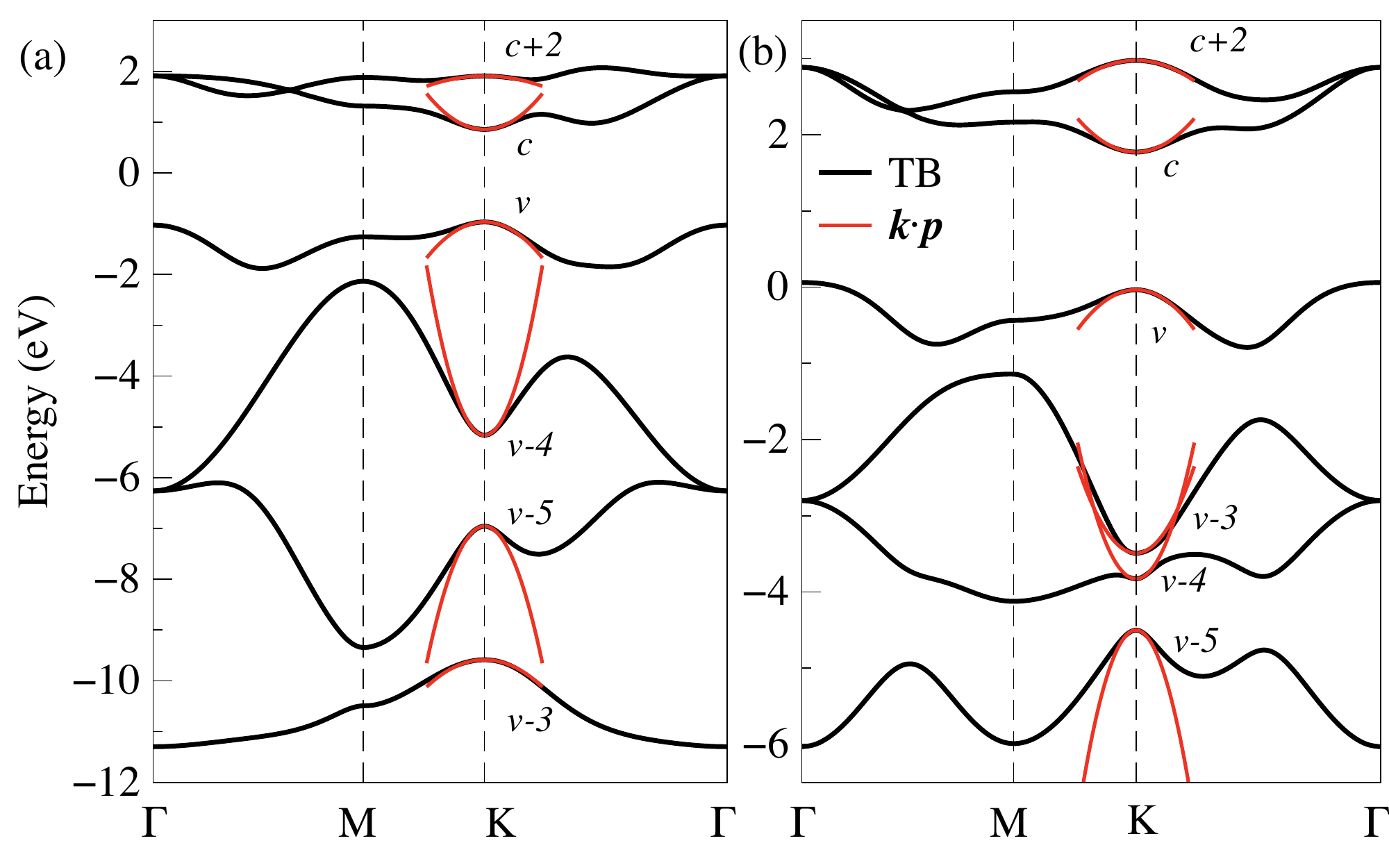}
\caption{\label{fig:fig1} Electronic spectra of MoS$_2$ calculated using TB models H. Rostami et al.~\cite{PhysRevB.92.195402} (a) and S. Fang et al.~\cite{PhysRevB.92.205108} (b). Red lines are \kp~quadratic dispersions at $\bm K$-point calculated using effective masses for each band (see text for details). Note the different order of deep valence bands in two panels.
}
\end{figure}

Figure~\ref{fig:fig1} shows the dispersion of electronic bands in MoS$_2$ calculated in the framework of TB models of Refs.~\cite{PhysRevB.92.195402, PhysRevB.92.205108}, respectively, in panels (b) and (a) and the dispersion $\eps_n = E_n + \hbar^2 q^2/2m_n$, where an effective mass of the $n$-th band is calculated similar to $n=c$ and $n=v$, see Eqs.~\eqref{eq:masses}, \eqref{eq:renormc}, \eqref{eq:renormv}. Figure~\ref{fig:fig1} illustrates that a certain care should be taken when labeling the valence bands: the bands are labeled according to its wave functions representations, see Tab.~\ref{tab:table1}, and the order might be different for different TB models. The effective masses $m_c$ and $m_v$ for different TB models are listed in Tab.~\ref{tab:gfactors}.

\begin{table*}[htpb]
\caption{\label{tab:gfactors} The values of $g$-factors and effective masses (in units of $m_0$) of the $v$ and $c$ bands calculated within the \kp~model based on different TB parameterizations (listed in footnotes).
}
\begin{ruledtabular}
\begin{tabular}{ccccccccc}
 & MoS$_2$~\footnote{TB model of Ref.~[\onlinecite{PhysRevB.92.195402}] for DFT calculations}& MoS$_2$~\footnote{TB model of Ref.~[\onlinecite{PhysRevB.92.205108}] for DFT calculations} &MoS$_2$~\footnote{TB model of Ref.~[\onlinecite{PhysRevB.92.205108}] for DFT+GW calculations} & MoS$_2$~\footnote{DFT+GW and TB model of this work (TB model based on Ref.~[\onlinecite{PhysRevB.92.195402}])} & MoS$_2$~\footnote{DFT+GW and TB model of this work (TB model based on Ref.~[\onlinecite{PhysRevB.92.205108}])}& MoSe$_2$~\footnote{TB model of Ref.~[\onlinecite{PhysRevB.92.205108}] for DFT calculations} & WS$_2$~\footnote{TB model of Ref.~[\onlinecite{PhysRevB.92.205108}] for DFT calculations} & WSe$_2$~\footnote{TB model of Ref.~[\onlinecite{PhysRevB.92.205108}] for DFT calculations}\\
 \hline

$m_v$ & -0.54 & -0.72 & -0.58 &-0.40 & -0.56 & -0.82 & -0.53 & -0.57 \\
$m_c$ & 0.54 & 0.86 & 0.90 &0.37 & 0.37 & 1.02 & 0.68 & 0.76\\
$g_v$ & 8.73 & 5.57 & 6.18 & 11.90 & 5.59 & 5.12 & 6.08 & 5.64 \\
$g_c$ & 7.82 & 5.41 & 6.83 & 10.15 & 1.77 & 5.12 & 6.13 & 5.79 \\
$g_{X^0} = g_c - g_v$ & -0.91 & -0.16 & 0.65 & -1.75 & -3.82 & 0 & 0.05 & 0.15
\end{tabular}
\end{ruledtabular}
\end{table*}

\section{Zeeman effect}

In this section we use the developed \kp~model to calculate the $g$-factors of electrons in conduction and valence bands. The main interest, however, is related to the exciton $g$-factor, which has been measured in a number of recent experiments by optical means. Single carrier Zeeman splittings can be determined, for instance, from the measurements of Shubnikov-de Haas oscillations.

 We consider the Zeeman splitting of electrons in $\bm K^\pm$~valleys in magnetic field $\bm B = (0, 0, B_z)$ directed normal to a monolayer plane. The Zeeman effect contains spin and valley contributions described by $g$-factors $g_0$ and $g_{\rm orb}$, respectively
\begin{equation}
\label{eq:Zeeman}
\mathcal H_B = \frac{g_0}{2} \mu_B B_z \sigma_z + \frac{g_{\rm orb}}{2} \mu_B B_z \tau_z\:.
\end{equation}
Here $\sigma_z$ is a spin operator ($\sigma_z = \pm 1$ for spin-up and spin-down electrons, respectively), and $\tau_z$ represents the valley degree of freedom ($\tau_z = \pm 1$ for $\bm K^+$ and $\bm K^-$ electrons, respectively), $\mu_B$ is the Bohr magneton. The effective $g$-factors of $\bm K^+$ and $\bm K^-$ electrons are defined as~\cite{wang2015}
\begin{equation}
\label{eq:gfactor}
g^{\bm K+}_{c, v} \equiv g_{c,v} = g_0 + g_{\rm orb}^{c,v}\:,\:\:\: g^{\bm K+}_{c, v} = -g^{\bm K-}_{c, v}\:.
\end{equation}

The valley term $g_{\rm orb}^{c,v}$ has an orbital nature and accounts for the mixing of the electronic bands by magnetic field. Within the \kp~scheme this mixing is obtained by replacing $\bm q$ in Eq.~\eqref{eq:H1} with $\bm q -(e/c\hbar) \bm A$, where $e = -|e|$ is the electron charge, and $\bm A$ is the vector potential of the magnetic field, resulting in (cf. Ref.~\cite{wang2015})
\begin{align}
\label{eq:gc}
g_{\rm orb}^c = \frac{4 m_0}{ \hbar^2}
 \left( - \frac{\gamma_5^2}{E_c - E_{v-3}} \right. & \left. + \frac{\gamma_3^2}{E_c - E_v} - \right. \\
\left. - \frac{\gamma_6^2}{E_c - E_{c+2}} \right. & \left.+ \frac{\delta_3^2}{E_c - E_{v-4}} \right)\:, \nonumber \\
\label{eq:gv}
g_{\rm orb}^v = \frac{4 m_0}{ \hbar^2}
 \left( \frac{\gamma_2^2}{E_v - E_{v-3}} \right. & \left. - \frac{\gamma_3^2}{E_v - E_c} + \right. \\
\left. + \frac{\gamma_4^2}{E_v - E_{c+2}} \right. & \left.  - \frac{\delta_4^2}{E_v - E_{v-5}} \right) \nonumber \:.
\end{align}
The spin $g$-factor $g_0$ comprises two contributions, namely, the bare electron $g$-factor ($g_0 = 2$) and a small contribution due to the spin-orbit interaction, which is not taken into account in our model. This contribution within the \kp~model is of the order of $\sim g_{\rm orb} \Delta_{\rm so}/\Delta E_{ij} \ll g_{\rm orb}$, where $\Delta_{\rm so}$ is the spin-orbit splitting of a given band, and $\Delta E_{ij}$ is a characteristic energy distance to other bands.

The $\sigma^+$ and $\sigma^-$ photoluminescence lines observed in experiment originate from the radiative recombination of neutral excitons $X^0$ with electrons occupying $\bm K^+$ and $\bm K^-$ valleys, respectively~\cite{PSSB:PSSB201552211}. Therefore the effective Zeeman splitting of $X^0$ is $\Delta_Z = g_{X^0} \mu_B B_z$ with~\cite{wang2015}
\begin{equation}
g_{X^0} = g_c - g_v\:.
\end{equation}
 In this difference, according to Eqs.~\eqref{eq:gfactor}, \eqref{eq:gc}, \eqref{eq:gv}, the contribution to $g_c$ and $g_v$, which occurs due to the mixing between $c$ and $v$ bands ($\propto \gamma_3^2$), cancels out, so that nonzero contributions to $g_{X^0}$ arise due to the mixing of $v$ and $c$ with deep valence and excited conduction bands.


\section{Landau levels}

The developed \kp~model allows for calculation of Landau levels in the system. For this purpose we make replacements $q_+ \to \sqrt{2} a^\dag/l_B$ and $q_- \to \sqrt{2} a/l_B$ in Eqs.~\eqref{eq:H1} and \eqref{eq:H2}, where $a^\dag$ and $a$ are the creation and annihilation operators acting in the space of Landau functions, $l_B = \sqrt{|e| \hbar/|B_z| c}$, and decompose the six-component wave function of the $j$-th band $\Psi^{(j)}$ in a series of the Landau level functions $\varphi_{n, q_y}$~\cite{rashba64}
\begin{equation}
\label{eq:wf_LL}
\Psi^{(j)} = \sum_{n \geq 0} 
\left(
\begin{array}{c}
a_n \\
b_n \\
c_n \\
d_n \\
e_n \\
f_n
\end{array}
\right)
\varphi_{n, q_y}\:,
\end{equation}
where $n$ and $q_y$ are the quantum numbers, and $a_n$, $b_n$, $c_n$, $d_n$, $e_n$, and $f_n$ are coefficients. The numeric diagonalization of obtained Hamiltonian yields the energy position of the $j$-th band in magnetic field $E_j(n, B_z)$. However this energy contains also the valley Zeeman term, described by the second term of Eq.~\eqref{eq:Zeeman}. To get rid of the Zeeman term we define Landau levels as $\eps_j(n, B_z) = 1/2 [E_j(n, B_z) + E_j(n, - B_z)]$. 

\begin{figure}[htpb]
\includegraphics[width=0.9\linewidth]{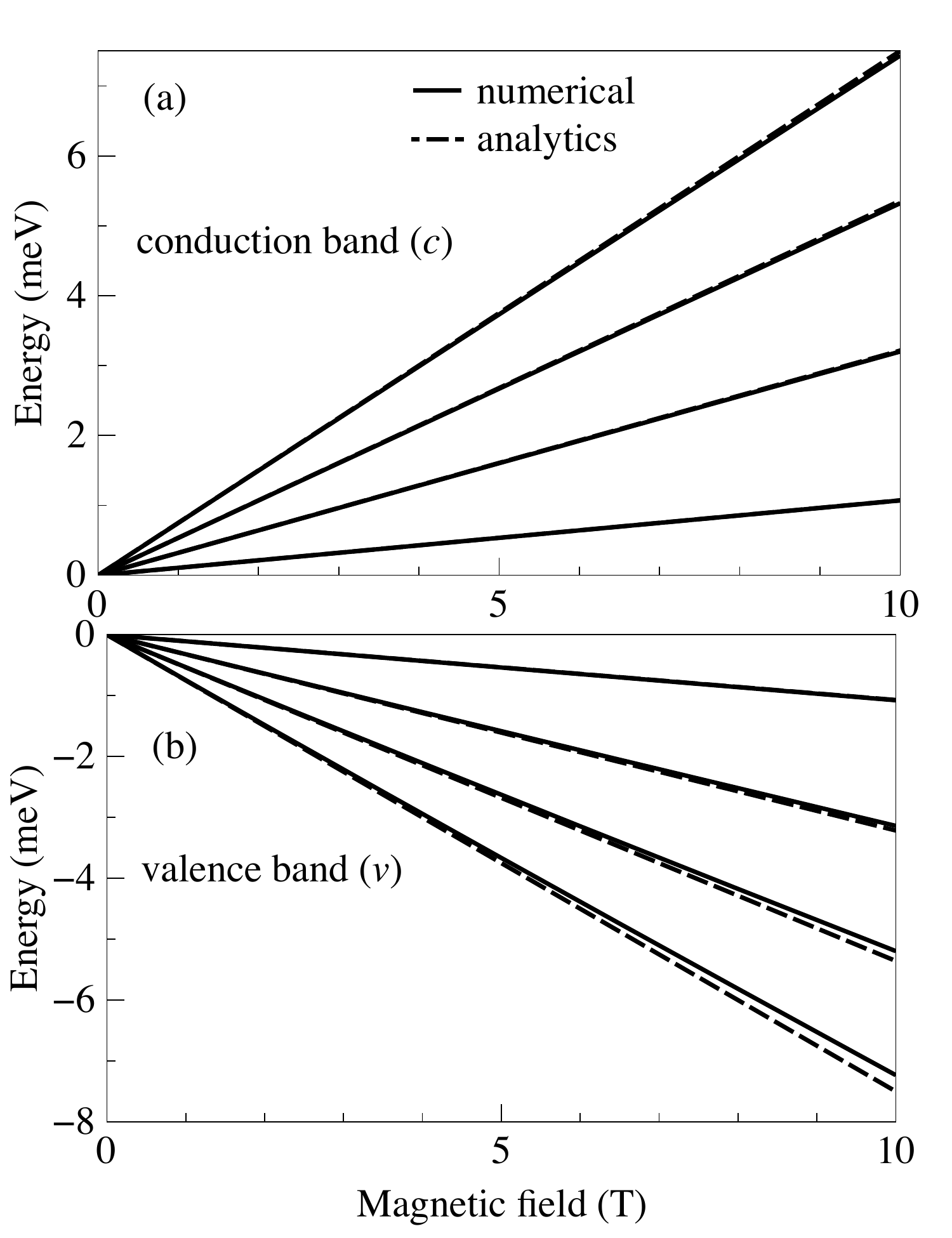}
\caption{\label{fig:fig2} Dispersion of the first four Landau levels in the bottom conduction ($c$) and topmost valence ($v$) bands calculated using parametrization (a) of the \kp~model (see Tab.~\ref{tab:table3}).  Solid lines show the results of numeric calculations using decomposition Eq.~\eqref{eq:wf_LL}, and dashed lines stand for a simple analytical formula $\eps_j(n, B_z) = \hbar \omega_c^{(j)} (n + 1/2)$ (see text for details).
}
\end{figure}

The dispersion of the first four Landau levels for $j = c$ and $j = v$ is presented in Fig.~\ref{fig:fig2}. 
For comparison we also show linear dispersions calculated using the simple formula $\eps_j(n, B_z) = \hbar \omega_c^{(j)} (n + 1/2)$, with $\omega_c^{(j)} = |e B_z|/m_j c$ and the effective masses $m_c$ and $m_v$ given by Eq.~\eqref{eq:masses}. One can see that the results given by the exact numeric diagonalization of the effective Hamiltonian and the simple analytical formula coincide in the wide range of $B_z$, the discrepancy in the valence band is more noticeable due to more pronounced band non parabolicity.

\section{Discussion}
Table~\ref{tab:gfactors} presents the values of $g$-factors and effective masses calculated within the developed \kp~model after Eqs.~\eqref{eq:gc}, \eqref{eq:gv} and Eqs.~\eqref{eq:masses}, \eqref{eq:renormc}, \eqref{eq:renormv}. Since the large contributions to $g_{c}$ and $g_v$ that originate from the mixing between $v$ and $c$ bands cancel out in the exciton $g$-factor, the value of $g_{X^0}$ is defined by the mixing with deep valence and excited conduction bands. The main contributions to $g_{X^0}$ in the studied parameterizations come from the mixing with $v-3$ and $c+2$ bands, i.e. from the terms $-\gamma_5^2/(E_c - E_{v-3})$ and $-\gamma_6^2/(E_c - E_{c+2})$ in Eq.~\eqref{eq:gc} and terms $\gamma_2^2/(E_v - E_{v-3})$ and $\gamma_4^2/(E_v - E_{c+2})$ in Eq.~\eqref{eq:gv}. As an example, the contribution from $c+2$ gives $\approx 15$~\% of the total $g_c$ value, and the contribution from $v-3$ gives $\approx 30$~\% of the total $g_v$ value for MoS$_2$ parametrization (a) in Tab.~\ref{tab:gfactors}. As seen from Tab.~\ref{tab:gfactors}, existing in literature TB models result in small absolute values of $g_{X^0}$, which are far from the experimentally measured $g_{X^0} \approx -4$.

Based on our own DFT and post-DFT ($GW$) calculations and TB fitting procedure (see computational details and dispersion of energy bands in Appendices~\ref{app:DFT}, \ref{app:TB}) we obtained two additional \kp~parameterizations for MoS$_2$, see columns (d) and (e) in Tabs.~\ref{tab:gfactors}, \ref{tab:table3}. For this purpose we used two different TB models, the eleven-parameters TB model of Ref.~[\onlinecite{PhysRevB.92.195402}] and the TB model of Ref.~[\onlinecite{PhysRevB.92.205108}] with twenty five independent parameters. 
Within the fitting procedure we numerically extracted parameters of the \kp~Hamiltonian and calculated exciton $g$-factor using Eqs.~\eqref{eq:gc}, \eqref{eq:gv}. We then used the value of $g_{X^0}$ as an extra fitting parameter (we fit it to the experimental value $g_{X^0} \approx -4$) additional to band dispersions and wave function coefficients.

The fitting procedure for the TB model of Ref.~[\onlinecite{PhysRevB.92.195402}] results in a good fit of the dispersion of $c$ and $v$ bands across high-symmetry paths of the Brillouin zone and only a slight change of energy position and dispersion of high conduction and deep valence bands compared to the original parameterization of Ref.~[\onlinecite{PhysRevB.92.195402}]. However this change is sufficient to obtain a large increase of $|g_{X^0}|$, $g_{X^0} \approx -1.75$ (see column (d) of Tab.~\ref{tab:gfactors}). Using the TB model of Ref.~[\onlinecite{PhysRevB.92.205108}] we were able to obtain $g_{X^0} \approx -3.82$ as well as a good fit for all six energy bands dispersions, wave function coefficients and effective masses, see column (e) in Tab.~\ref{tab:gfactors}.

The wide spread of calculated $g_{X^0}$ values underlies the sensitivity of $g_{X^0}$ to a given parametrization of a DFT or a TB model. Hence, along with effective masses and energy gaps the value of $g_{X^0}$ may serve as a test tool for a given parametrization of a DFT or a TB model. The TB parameterizations that fit $g_{X^0}$ can be obtained in principle for other materials (e.g. MoSe$_2$, WS$_2$, and WSe$_2$), however this is beyond the scope of the present work.

\begin{figure}[htpb]
\includegraphics[width=0.9\linewidth]{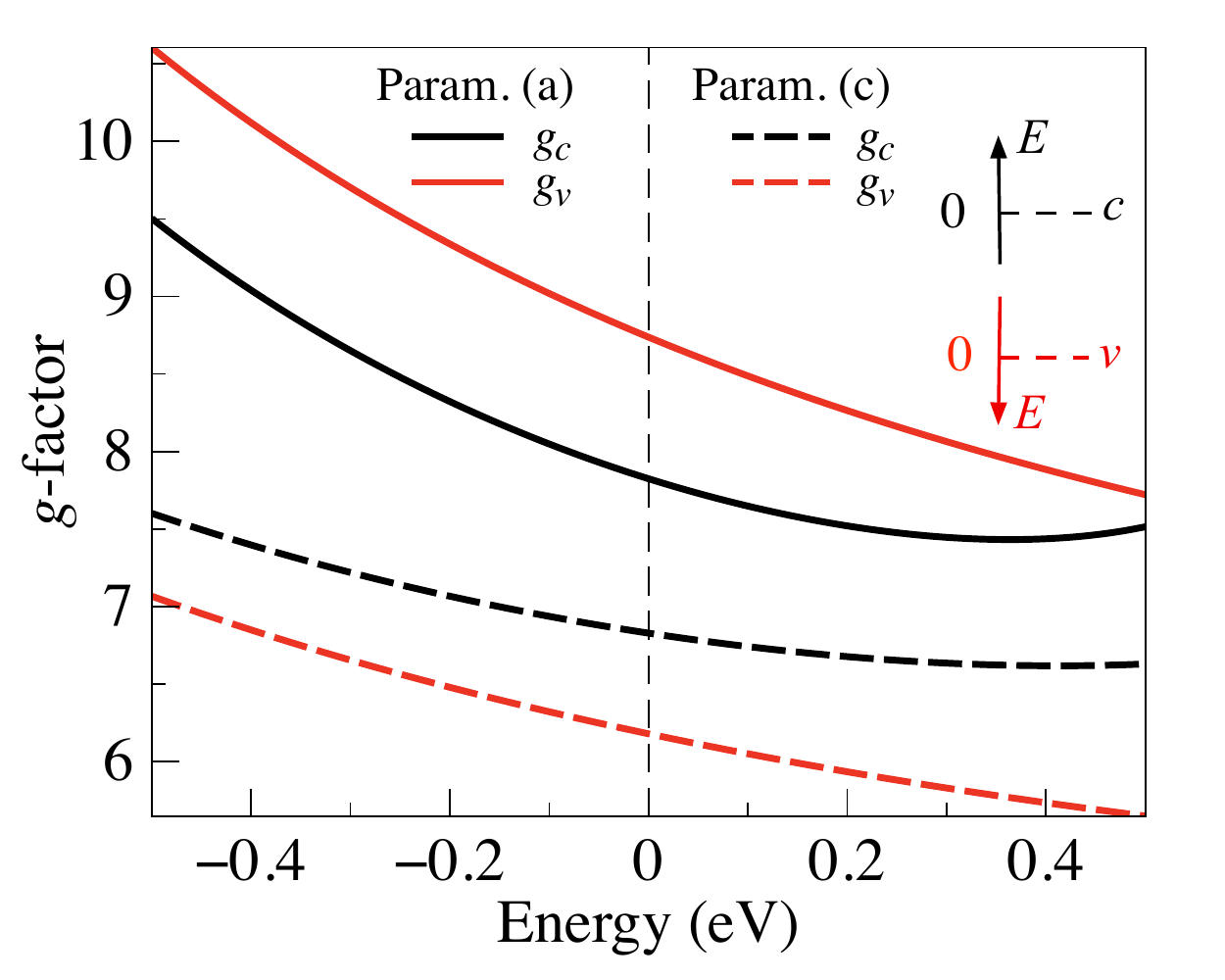}
\caption{\label{fig:fig3}
Conduction and valence band $g$-factors, $g_c$ and $g_v$, as functions of electron and hole energy, respectively. The energies are counted from the bottom of the conduction band and from the top of the valence band as shown in the inset. The solid and dashed lines show results of calculations for parameterizations (a) and (c) from Tab.~\ref{tab:table3}, respectively.
}
\end{figure}

So far in our theory we treated electron and hole in the exciton independently, neglecting the Coulomb interaction between charge carriers. It is well known, however, that the exciton binding is large in TMDCs and plays a significant role in optical experiments~\cite{PhysRevLett.113.026803, PhysRevLett.113.076802}. The effects of Coulomb interaction between an electron and a hole as well as localization by an in-plane potential~\cite{:/content/aip/journal/apl/108/14/10.1063/1.4945268} can be estimated by introducing the $g$-factor dependence on a charge carrier energy. Such a dependence is derived by simply replacing $E_c$ with $E_c + \Delta E_c$ in Eq.~\eqref{eq:gc} and $E_v$ with $E_v - \Delta E_v$ in Eq.~\eqref{eq:gv}, where $\Delta E_c$ and $\Delta E_v$ are the energy shifts of electron and hole levels~\cite{ivchenko05a} due to either localization or Coulomb binding. This dependence is depicted in Fig.~\ref{fig:fig3}. Note that negative energies $\Delta E_c$, $\Delta E_v$ reflect the binding of electron and hole in an exciton, whereas positive $\Delta E_c$, $\Delta E_v$ correspond to localization of a charge carrier in a quantum dot. One can see that within a typical scale of exciton binding energy in TMDCs, $E_B \sim 400$~meV, $g_c$ and $g_v$ change significantly, resulting in a possible enhancement $\Delta|g_{X^0}| \approx 1$.

\section{Conclusions}

To conclude, we developed a six-band \kp~model that describes the electronic states of monolayer TMDCs in $\bm K$-valleys. The set of parameters for the \kp~model is uniquely determined
 by decomposing eleven-band tight-binding models in the vicinity of $\bm K^\pm$-points. Using existing in literature TB models we were able to derive systematic parametrizations for different materials (MoS$_2$, WS$_2$, MoSe$_2$, WSe$_2$) and different TB Hamiltonians. Using the derived six-band \kp~Hamiltonian we calculated effective masses, Landau levels, and the effective exciton $g$-factor $g_{X^0}$ in different TMDCs. 
 We showed that the main contributions to $g_{X^0}$ result from the mixing with excited conduction band $c+2$ and deep valence band $v-3$. We also obtained two additional sets of \kp~parameters by developing our own TB parameterizations based on simultaneous fitting of ab-initio calculated energy dispersion and the value of $g_{X^0}$. 

 The \kp~parameterizations extracted from existing TB models result in small absolute values of $g_{X^0}$, which are far from the experimentally measured $g_{X^0} \approx -4$. However as we showed using our additional sets of \kp~parameters, the change in parameters, which only slightly affects the dispersion of higher conduction and deep valence bands, may result in a significant increase of $|g_{X^0}|$. As a result, we obtained $g_{X^0} \approx -1.75$ and $g_{X^0} \approx -3.82$ for the two sets. Such a high parameter sensitivity of $g_{X^0}$ opens a way to further improvement of DFT and TB models, since $g$-factor modeling requires at the same time an accurate description of deep valence and high conduction bands.

\acknowledgments
I.C.G. thanks the CNRS and the ANR MoS2ValleyControl project for financial support. He also acknowledges the CALMIP initiative for the generous allocation of computational times, through Project p0812, as well as the GENCI-CINES, GENCI-IDRIS, and GENCI- CCRT for Grant No. x2016096649. 

M.V.D. was partially supported by RFBR projects No. 14-02-00168 � and No. 16-32-60175, the Russian Federation President Grant No. MK-7389.2016.2 and the Dynasty foundation. M.V.D. thanks M. M. Glazov for fruitful discussions.

\appendix

\section{Parameters of $kp$~model} \label{app:kp}

Parameters of the \kp~Hamiltonians~\eqref{eq:H1} and \eqref{eq:H2} for different parametrizations, materials and tight-binding models are listed in Tab.~\ref{tab:table3}.

\begin{table*}[hbtp]
\caption{\label{tab:table3} Parameters of the \kp~model as introduced in Eqs.~\eqref{eq:H1}, \eqref{eq:H2}. The values of $\gamma_i$ and $\delta_i$ are given in eV\AA, the units of $E_i$ are eV, the effective masses $m'_{n}$ are given in the units of $m_0$. Parameterizations are based on TB models listed in footnotes.}
\begin{ruledtabular}
\begin{tabular}{ccccccccc}
  & MoS$_2$~\footnote{TB model of Ref.~[\onlinecite{PhysRevB.92.195402}] for DFT calculations}& MoS$_2$~\footnote{TB model of Ref.~[\onlinecite{PhysRevB.92.205108}] for DFT calculations} &MoS$_2$~\footnote{TB model of Ref.~[\onlinecite{PhysRevB.92.205108}] for DFT+GW calculations} & MoS$_2$~\footnote{DFT+GW and TB parametrization of this work (TB model based on Ref.~[\onlinecite{PhysRevB.92.195402}])} & MoS$_2$~\footnote{DFT+GW and TB parametrization of this work (TB model based on Ref.~[\onlinecite{PhysRevB.92.205108}])}& MoSe$_2$~\footnote{TB model of Ref.~[\onlinecite{PhysRevB.92.205108}] for DFT calculations} & WS$_2$~\footnote{TB model of Ref.~[\onlinecite{PhysRevB.92.205108}] for DFT calculations} & WSe$_2$~\footnote{TB model of Ref.~[\onlinecite{PhysRevB.92.205108}] for DFT calculations}\\
 \hline
$E_{v-5}$ & -6.96 & -4.50 & -4.99 & -6.88 & -5.20 & -4.42 & -5.27 & -5.14 \\ 
$E_{v-4}$ & -5.17 & -3.83 & -4.32 & -4.15 & -4.66 & -3.70 & -4.21 & -4.02 \\ 
$E_{v-3}$ & -9.59 & -3.49 & -3.62 & -10.52 & -4.18 &-3.36 & -3.82 & -3.67 \\ 
$E_{v}$ & -0.97 & -0.03 &  0 & 0 & -0.05 & -0.05 & 0.04 & 0.02 \\ 
$E_{c}$ & 0.86 & 1.77 & 2.48 & 2.47 & 2.44 & 1.52 & 2.00 & 1.69 \\ 
$E_{c+2}$ & 1.91 & 2.98 & 4.04 &3.96 & 4.60 & 2.50 & 3.36 & 2.80 \\ 
\hline
\hline
$\gamma_2$ & -5.75 & 1.62 & 2.08 & -8.00 & -0.88 & 1.50 & 1.62 & 1.49 \\ 
$\gamma_3$ & 4.27 & 3.39 & 4.43 & 5.93 & 4.65 & 2.96 & 3.91 & 3.43 \\
$\gamma_4$ & -0.87 & -0.92 & -2.14 &-1.77 & -3.05 & -0.91 & -1.53 & -1.44 \\
$\gamma_5$ & 2.57 & -2.66 & -3.07 & 3.36 & -8.27 & -2.44 & -3.26 & -3.04 \\
$\gamma_6$ & 1.33 & 0.94 & 1.52 & 1.79 & 0.67 & 0.84 & 1.21 & 1.05 \\
\hline
\hline
$\delta_1$ & 3.19 & -4.20 & -5.14 &4.05 & -3.80 & -3.86 & -4.95 & -4.52 \\
$\delta_2$ & 0.80 & -0.19 & -0.50 &1.26 & 3.55 & -0.16 & -0.30 & -0.29 \\
$\delta_3$ & -0.61 & 2.08 & 2.53 &0.55 & -2.63 & 2.11 & 2.23 & 2.25 \\
$\delta_4$ & -2.05 & 0.14 & 0.02 &-2.09 & -0.26 & -0.06 & 0.18 & -0.06 \\
$\delta_5$ & 1.74 & 2.06 & 2.15 &2.28 & -0.42 & 1.79 & 2.15 & 1.88 \\
$\delta_6$ & 1.45 & 0.69 & 0.69 &2.23 & -0.23 & 0.48 & 0.32 & 0.07 \\
$\delta_7$ & 7.49 & 4.45 & 5.05 &6.53 & 3.90 & 4.81 & 4.78 & 5.14 \\
\hline
\hline
$m'_{v-5}$ & 0.87 & 0.76 & 0.67 & 0.85 & 0.44 & 0.67 & 0.64 & 0.57 \\
$m'_{v-4}$ & 1.34 & 0.83 & 0.71 & 2.00 & 1.22 & 0.78 & 0.84 & 0.80 \\
$m'_{v-3}$ & 6.09 & 6.92 & 14.00 & 1.64 & 0.62 & 7.69 & 9.50 & 12.32 \\
$m'_{v}$ & -2.81 & 6.37 & 3.04 & -3.39 & 1.03 & 6.58 & 6.64 & 7.16 \\
$m'_{c}$ & -1.96 & -1.16 & -0.90 & -1.33 & -0.40 & -1.18 & -1.02 & -1.04 \\
$m'_{c+2}$ & -0.70 & -0.60 & -0.47 & -0.59 & -0.36 & -0.63 & -0.53 & -0.55 
\end{tabular}
\end{ruledtabular}
\end{table*}

\section{Computational details of DFT+GW method} \label{app:DFT}

The atomic structures and the quasi-particle band structures have been obtained from DFT calculations using the VASP 
package \cite{Kresse:1993a,kresse:prb:96}. The Perdew-Burke-Ernzerhof (PBE)~\cite{Perdew:1996prl} functional was used as approximation of the exchange-correlation electronic term. The software uses the plane-augmented wave scheme~\cite{blochl:prb:94,kresse:prb:99} to treat core electrons. 
Fourteen electrons for Mo, W atoms and six for S, Se ones are explicitly included in the valence states. All atoms are allowed to relax with a force convergence criterion below
 $0.005$ eV/\AA. A grid of 12$\times$12$\times$1 $k$-points has been used, in conjunction with a vacuum height of 17 \AA, to take benefit of error's cancellation in the band gap estimates~\cite{Huser:2013a}. 
A gaussian smearing with a width of 0.05 eV was used for partial occupancies, when a tight electronic minimization tolerance  
of $10^{-8}$ eV was set to determine with a good precision the corresponding derivative of the orbitals with respect to $k$ needed 
in quasi-particle band structure calculations. Spin-orbit coupling was not included to determine eigenvalues and wave functions as input for the full-frequency-dependent $GW$ 
calculations~\cite{Shishkin:2006a} performed at the $G_0W_0$ level. The total number of states included in the $GW$ procedure was set to 600, after a careful check of the direct band gap convergence, to be smaller than 0.1 eV.

\section{Details of TB fitting procedure and additional TB parameterizations} \label{app:TB}

Most of the modern TB parametrizations are made to reproduce the energy bands of ab-initio calculations. The parameter set is usually found by minimizing the function:
\begin{equation}
\label{eq:TBfit}
f(\{t_i\}) = \sum \limits_{i, \bm k} w_{i, \bm k} \left( E_{i, \bm k}^{\rm TB} - E_{i, \bm k}^{GW} \right)^2\:,
\end{equation}
where $\{t_i\}$ are the TB-parameters, $i$ and $\bm k$ denote the number of the electronic band and the wave vector, respectively, and $w_{i,\bm k}$ are the weight coefficients. In Eq.~\eqref{eq:TBfit} $E^{\rm TB}$ are the tight binding energies, which depend on the particular TB Hamiltonian and parameters used, and $E^{GW}$ are the starting-point values, obtained by DFT+$GW$ or another ab-initio method. 
We found, however, that even if the parameter set reproduces the electronic bands with great accuracy, it does not necessarily give satisfying values of the $g$-factor. In order to overcome this problem, we included the calculation of the $g$-factor in our optimization procedure and varied the TB parameters to fit both the $GW$-energies and $g$-factor values. To prevent the order change of the energy bands during optimization we also included the error in the eigenvectors at the $K$-point. The weights were concentrated in the vicinity of the $K$ and $\Gamma$ points of the hexagonal Brillouin zone and had higher values for $v$ and $c$ bands for better reproduction of the most important electronic states. The fitting was carried out by an adaptive random search algorithm until a compromise between the quality of the band structure and $g$-factor value for MoS$_2$ was found.

\begin{figure}[htpb]
\includegraphics[width=0.47\textwidth]{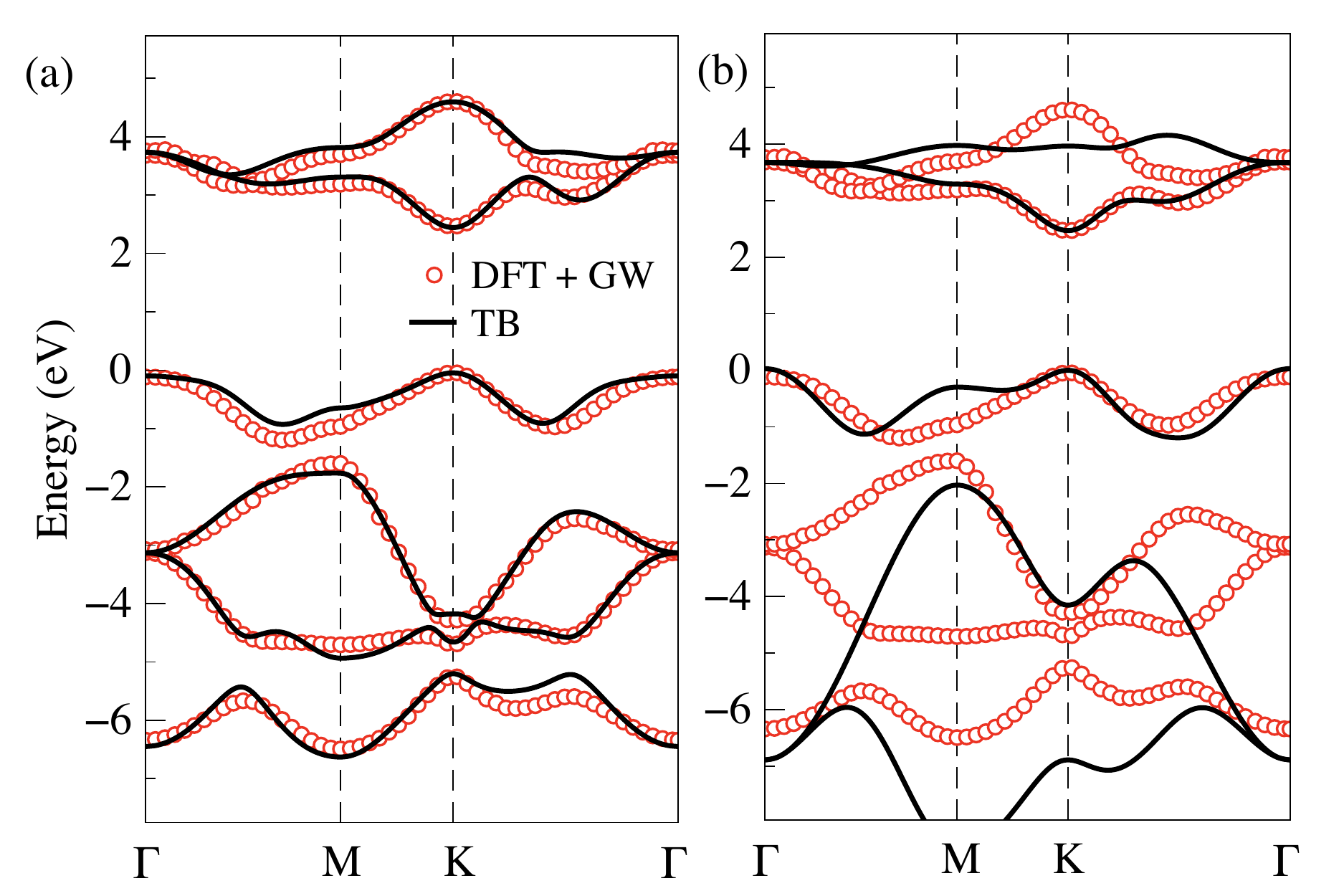}
\caption{\label{fig:DFT_TB} DGT+GW and TB calculations for electronic energy dispersion in MoS$_2$. (a) The fit of TB model of Ref.~[\onlinecite{PhysRevB.92.205108}], (b) the fit of TB model of Ref.~[\onlinecite{PhysRevB.92.195402}]. The zero energy was aligned to the top of the valence band $v$.
}
\end{figure}

Resulting parameterizations of TB Hamiltonians of Refs.~[\onlinecite{PhysRevB.92.205108}] and [\onlinecite{PhysRevB.92.195402}] are presented in Tabs.~\ref{tab:table4} and \ref{tab:table5}. The resulting energy dispersions in comparison with DFT+GW calculations are presented in Fig.~\ref{fig:DFT_TB}. Based on these two sets of TB parameters we obtained \kp~parameterizations listed in Tab.~\ref{tab:table3}, columns (d) and (e), with effective masses and $g$-factor listed in Tab.~\ref{tab:gfactors}, columns (d) and (e).

\begin{table}[htpb]
\caption{\label{tab:table4} Parameters of the TB Hamiltonian (in units of eV) of Ref.~[\onlinecite{PhysRevB.92.205108}] for MoS$_2$ obtained after fitting of $GW$ calculations.}
\begin{ruledtabular}
\begin{tabular}{ccccccc}
 $\epsilon_6$ & $\epsilon_7 = \epsilon_8$ & $\epsilon_9$ & $\epsilon_{10} = \epsilon_{11}$ & $t_{6,6}^{(1)}$ & $t_{7,7}^{(1)}$ & $t_{8,8}^{(1)}$  \\
  -0.913  & 0.251 & -1.538 & -2.264 & -0.922 & 0.437 & -0.668\\
\hline
 $t_{9,9}^{(1)}$ & $t_{10,10}^{(1)}$ & $t_{11,11}^{(1)}$ & $t_{6,8}^{(1)}$ & $t_{9,11}^{(1)}$ & $t_{6,7}^{(1)}$ & $t_{7,8}^{(1)}$ \\
 0.240 & 1.106 & -0.003 & 0.046 & -0.041 & -0.762 & -0.400 \\
 \hline
$t_{9,10}^{(1)}$ & $t_{10,11}^{(1)}$ & $t_{9,6}^{(5)}$ & $t_{11,6}^{(5)}$ & $t_{10,7}^{(5)}$ & $t_{9,8}^{(5)}$ & $t_{11,8}^{(5)}$ \\
-0.168 & -0.133 & -0.975 & 0.016 & 1.829 & 0.914 & -0.045 \\
\hline
$t_{9,6}^{(6)}$ & $t_{11,6}^{(6)}$ & $t_{9,8}^{(6)}$ & $t_{11,8}^{(6)}$ \\
0.935  & 0.945  & 0.796  & 0.449
\end{tabular}
\end{ruledtabular}
\end{table}

\begin{table}[htpb]
\caption{\label{tab:table5} Parameters of the TB Hamiltonian (in units of eV) of Ref.~[\onlinecite{PhysRevB.92.195402}] for MoS$_2$ obtained after fitting of $GW$ calculations.}
\begin{ruledtabular}
\begin{tabular}{ccccccc}
 $\epsilon_0$ & $\epsilon_2$ & $\epsilon_p$ & $\epsilon_z$ & $V_{pd\sigma}$ & $V_{pd\pi}$ & $V_{dd\sigma}$  \\
  -5.707  & -5.784 & -8.319 & -12.171 & 4.791 & -1.606 & -1.221\\
\hline
 $V_{dd\pi}$ & $V_{dd\sigma}$ & $V_{pp\sigma}$ & $V_{pp\pi}$\\
 0.526 & 0.359 & 0.905 & -0.396  \\
 \end{tabular}
\end{ruledtabular}
\end{table}

\normalem

%

\end{document}